\documentclass[aps,prd,amsmath,amssymb,showpacs]{revtex4}

\usepackage{epsfig,float}
\usepackage{graphicx}
\usepackage{dcolumn}
\usepackage{morefloats}
\usepackage{color}
\usepackage{slashed}
\usepackage{bbm}

\newcommand{\Lag}{\mathcal{L}}
\newcommand{\Amp}{\mathcal{A}}

\newcommand{\intlf}{\int\!\frac{\mathrm d^4 l}{(2\pi)^4}}

\newcommand{\intlz}{\int^\infty_0\!\mathrm d l}

\newcommand{\eps}{\epsilon}

\newcommand{\lt}{\left}
\newcommand{\rt}{\right}

\newcommand{\Lam}{\Lambda}

\newcommand{\gam}{\gamma}

\newcommand{\fr}{\frac}
\newcommand{\sq}{\sqrt}
\newcommand{\cd}{\cdot}

\newcommand{\mr}{\mathrm}

\newcommand{\be}{\begin{equation}}
\newcommand{\ee}{\end{equation}}
\newcommand{\ba}{\begin{eqnarray}}
\newcommand{\ea}{\end{eqnarray}}
\newcommand{\nn}{\nonumber}

\newcommand{\ffn}{\mathrm{exp}(-2l^2/\Lambda^2)}

\newcommand{\fc}{\frac}
\newcommand{\st}{\sqrt}

\begin{document}

\title{The impact of $S$-wave thresholds $D_{s1}\bar{D}_{s}+c.c.$ and $D_{s0}\bar{D}^*_{s}+c.c.$ on vector charmonium spectrum}

\author{Zheng Cao$^{1,2}$\footnote{{\it Email address:} caoz@ihep.ac.cn}
 and Qiang Zhao$^{1,2,3}$\footnote{{\it Email address:} zhaoq@ihep.ac.cn} }

\affiliation{$^1$ Institute of High Energy Physics and Theoretical Physics Center for Science Facilities,
        Chinese Academy of Sciences, Beijing 100049, China}

\affiliation{$^2$  School of Physical Sciences, University of Chinese Academy of Sciences, Beijing 100049, China}

\affiliation{$^3$ Synergetic Innovation Center for Quantum Effects and Applications (SICQEA), Hunan Normal
University, Changsha 410081, China}

\begin{abstract}
By studying the very closely lied $D_{s1}\bar{D}_{s}+c.c.$ and $D_{s0}\bar{D}^*_{s}+c.c.$ thresholds at about 4.43 GeV we investigate the possibilities of a sizeable molecular component inside the $\psi(4415)$  in a unitarization scheme. The $\psi(4415)$ is conventionally assigned as the $\psi(4S)$ state in the quark model. We also consider possible mixings between the $\psi(4S)$ and the quark model state $\psi(2D)$ given their $S$-wave couplings to both states. A significant coupling for $\psi(4415)$ to the nearby $S$-wave $D_{s1}\bar{D}_{s}+c.c.$ and $D_{s0}\bar{D}^*_{s}+c.c.$ thresholds may lead to formation of exotic states $Z_{cs}$ in the decay of $\psi(4415)\to J/\psi K\bar{K}$, which is the strange partner of $Z_c(3900)$, but has flavor contents of $c\bar{c}q\bar{s}$ (or $c\bar{c}s\bar{q}$) with $q$ denoting $u/d$ quarks. Similar to the production of $Z_c(3900)$ in $e^+e^-\to Y(4260)\to J/\psi\pi\pi$ via the intermediate $D_1(2420)\bar{D}+c.c.$ where the so-called ``triangle singularity (TS)" kinematics can be recognized, we find that the process $e^+e^-\to \psi(4415)\to J/\psi K\bar{K}$ via the intermediate $D_{s1}\bar{D}_{s}+c.c.$ and $D_{s0}\bar{D}^*_{s}+c.c.$ is located at the edge of the TS kinematics. It can provide an ideal case for examining the TS effects and make a difference between a genuine pole structure and kinematic enhancement.

\end{abstract}
\date{\today}

\maketitle
\section{Introduction}\label{sec:introduction}
During the past decade the observations of a large number of hadronic exotic candidates have initiated tremendous activities and efforts on understanding their dynamic nature in both experiment and theory. Most of these heavily flavored states which are tentatively named by ``$XYZ$" are intimately related to some nearby $S$-wave thresholds. This seems to provide important clues for understanding their intrinsic structures. Typical examples include $X(3872)$ and $Z_c(3900)$~\cite{Ablikim:2013mio} which are close to the $D\bar D^*+c.c.$ threshold, and  $Z_c(4020)$~\cite{Ablikim:2013wzq} to the $D^*\bar D^*$ threshold. Their bottomed correspondences are $Z_b(10610)$ and $Z_b(10650)$~\cite{Belle:2011aa} which are located at the $B\bar{B}^*+c.c.$ and $B^*\bar{B}^*$ thresholds, respectively. In the vector charmonium spectrum the mysterious $Y(4260)$ seems to be closely related to the $S$-wave $D_1(2420)\bar{D}+c.c.$ threshold in order to understand many new experimental observations of its exclusive decays. Recent studies indicate strong evidence for the hadronic molecule component of $D_1(2420)\bar{D}+c.c.$ in its wavefunction while a compact core should also be present as the consequence of heavy quark spin symmetry (HQSS) breaking effects~\cite{Wang:2013cya,Wang:2013kra,Cleven:2013mka,Qin:2016spb,Xue:2017xpu}.

Following these interesting discoveries, many theoretical interpretations are proposed in the literature to understand the internal structures of those exotic candidates. One can refer to several recent review articles, Refs.~\cite{Guo:2017jvc,Chen:2016spr,Olsen:2014qna,Lebed:2016hpi,Esposito:2016noz} and references therein, for the up-to-date status of the theoretical and experimental progresses. Some of the broadly discussed scenarios, which were motivated by the early observations of charmonium exotic candidates $X(3872)$ and $Y(4260)$, include hadro-charmonia~\cite{Voloshin:2007dx,Dubynskiy:2008mq}, tetraquarks~\cite{Maiani:2004vq}, loosely bound molecules~\cite{Close:2003sg,Thomas:2008ja}, and hybrids~\cite{Close:2005iz,Zhu:2005hp,Kou:2005gt}. Specific kinematic effects were also discussed in the literature as an interpretation of some of the threshold phenomena.  In Refs.~\cite{Bugg:2011jr,Chen:2011pv,Swanson:2015bsa} it was proposed that CUSP effects from two-body unitarity cut can account for some of these enhancements near threshold. However, in Ref.~\cite{Guo:2014iya}, it was demonstrated that although the CUSP effects can result in some structures, it is still not possible to produce pronounced and narrow near-threshold peaks without introducing physical poles. In contrast, another kinematic effect, i.e. the triangle singularity (TS), which is the leading singularity of the triangle loop transition amplitude, can possibly produce observable threshold enhancements~\cite{Wang:2013cya,Wang:2013hga,Liu:2013vfa,Liu:2015taa,Liu:2015cah,Szczepaniak:2015eza}. Special features arising from such a mechanism have been investigated in different processes~\cite{Guo:2015umn,Liu:2015fea,Wang:2013cya,Wu:2011yx,Guo:2014qra,Bayar:2016ftu,Wang:2016dtb}, and have attracted a lot of attention from the community. Although in some cases a pole contribution is needed in order to better describe the lineshapes~\cite{Cleven:2013mka,Qin:2016spb}, the TS mechanism turns out to be crucial for a better understanding of the threshold phenomena~\cite{Liu:2015taa,Guo:2017jvc}. In Refs.~\cite{Hanhart:2015cua,Guo:2016bjq}, a practical parametrization for the lineshapes of the near-threshold states is proposed. Based on the Lippmann-Schwinger equations for the coupled channel problem, this approach incorporates the inelastic channels additively with the unitarity and analyticity constraints for the $t$ matrix.

Motivated by the interest of searching for exotic hadrons and the special role played by the TS, in this work we investigate the very closely lied $D_{s1}\bar{D}_{s}+c.c.$ and $D_{s0}\bar{D}^*_{s}+c.c.$  thresholds which are the lowest $S$-wave charmed-strange open thresholds in the vector charmonium spectrum. For the convenience we note these two thresholds by $D_{s1}\bar{D}_{s}$ and $D_{s0}\bar{D}^*_{s}$ as follows in this work. These two thresholds are located close to some of the quark model states, in particular, the $\psi(4415)$, which is assigned as the $\psi(4S)$ state. At this moment, the experimental information on the properties of higher charmonium states is still very limited.
In 2007, Belle Collaboration investigated the $J/\psi K^+K^-$ final states in $e^+e^-$ annihilations via the initial-state radiation (ISR) from threshold to the center of mass (c.m.) energy of 6.0 GeV~\cite{Yuan:2007bt}. The measured cross sections seemed to indicate the need for the $\psi(4415)$. However, the limited statistics did not allow conclusions either on the detailed properties of $\psi(4415)$, or its possible correlations with the $S$-wave $D_{s1}\bar{D}_{s}$ and $D_{s0}\bar{D}^*_{s}$ thresholds. Recently, the BESIII Collaboration publish the analysis of $e^+e^-\to J/\psi K\bar{K}$ with the center of mass (c.m.) energy from 4.189 to 4.600 GeV where a clear and broad structure around 4.45$\sim$4.46 GeV is observed~\cite{Ablikim:2018epj}. This structure has a peak position apparently higher than the $\psi(4415)$. This improved experimental result may provide a better understanding of the $\psi(4415)$ and its relation with the nearby $D_{s1}\bar{D}_{s}+c.c.$ and $D_{s0}\bar{D}^*_{s}+c.c.$  thresholds.

Actually, given a strong enough $S$-wave coupling for an open threshold to the nearby quark model state, it may result in mixings between the quark model state and a molecular state which is dynamically generated by the open channel through unitarization. This is similar to what have been investigated in various channels correlated with the $D_1(2420)\bar{D}$ threshold in the interpretation of the $Y(4260)$~\cite{Wang:2013cya,Wang:2013kra,Cleven:2013mka,Qin:2016spb,Xue:2017xpu}.

Apart from the $\psi(4415)$, which is commonly assigned as the $\psi(4S)$ state in the quark model, the other nearby quark model states include the $\psi(4160)$ as the $\psi(2D)$ state below the $D_{s1}\bar{D}_{s}$ and $D_{s0}\bar{D}^*_{s}$ thresholds, and the $\psi(3D)$ which is supposed to be above these two thresholds~\footnote{We do not consider the exotic candidate $Y(4260)$ in this analysis although as the candidate for the $D_1(2420)\bar{D}+c.c.$ hadronic molecule, it may have some dynamical correlations with the $D_{s1}D_{s}$ and $D_{s0}D^*_{s}$  thresholds as a subleading effect. We note that the new data from BESIII~\cite{Ablikim:2018epj} show a narrow structure around 4.2 GeV in $J/\psi K\bar{K}$. In contrast, the data around $4.4\sim 4.5$ GeV indicate a broad enhancement.}. The $\psi(3D)$ was predicted to have a mass around 4.52 GeV in the quark model~\cite{Godfrey:1985xj}, and in Ref.~\cite{vanBeveren:2010mg} a mass region of $4.53\sim 4.58$ GeV is extracted which seems to be consistent with the quark model expectation. An interesting feature arising from the mass ordering of the quark model states in this energy region is that $m_{\psi(2D)}<m_{\psi(4S)}<m_{\psi(3D)}$~\cite{Godfrey:1985xj}. Given that the production of the $D$-wave states will be suppressed in the HQSS limit in $e^+e^-$ annihilations, we anticipate that the nearby $S$-wave thresholds should have larger impact on the $S$-wave states instead of the $D$-wave states. We take this as a reasonable approximation in the analysis. Moreover, since the masses of the $\psi(2D)$ and $\psi(3D)$ are relatively far away from the $D_{s1}\bar{D}_{s}$ and $D_{s0}\bar{D}^*_{s}$ thresholds, it will further justify that the open threshold effects on the $\psi(2D)$ and $\psi(3D)$ states should be suppressed. As a test of this expectation, we include the better established $\psi(4160)$ in the analysis and examine the impact of the $D_{s1}\bar{D}_{s}$ and $D_{s0}\bar{D}^*_{s}$ thresholds on its properties.

Similar to the production process of $Z_c(3900)$ in $e^+e^-\to Y(4260)\to J/\psi\pi\pi$ where the $S$-wave threshold  $D_1(2420)\bar{D}+c.c.$ plays a crucial role for understanding the properties of $Y(4260)$ and $Z_c(3900)$~\cite{Wang:2013cya}, the process $e^+e^-\to J/\psi K\bar{K}$ around the mass region of the thresholds of $D_{s1}\bar{D}_{s}$ and $D_{s0}\bar{D}^*_{s}$ in association with $\psi(4415)$ may also provide important clues for the production of exotic state $Z_{cs}$, which is the strange partner of $Z_c(3900)$, but has flavor contents of $c\bar{c}q\bar{s}$ (or $c\bar{c}s\bar{q}$) with $q$ denoting $u/d$ quarks. In particular, we are interested in the role played by the TS mechanism, which can be accessed in this channel. In Ref.~\cite{Liu:2015cah}, the TS mechanism corresponding to similar charmed-strange meson thresholds but with final states of $J/\psi$ and a hidden $s\bar s$ is also investigated. To be more specific, given that the initial vector states can first couple to $D_{s1}\bar{D}_{s}$ or $D_{s0}\bar{D}^*_{s}$, the intermediate $D_{s1}$ or $D_{s0}$ can then rescatter against $\bar{D}_s$ or $\bar{D}_s^*$ by exchanging $D^*$ or $D$, respectively, before converting into a Kaon, and then the interactions between the exchanged $D^*$ (or $D$) and $\bar{D}_s$ (or $\bar{D}_s^*$) will form $J/\psi$ and an anti-Kaon. Such a transition is via a triangle diagram, and for specific kinematics all these three internal particles may approach their on-shell conditions simultaneously. Such a kinematic condition is called the TS condition and it brings the leading singular amplitude to the loops. Actually, around the mass region of $\psi(4415)$, the kinematics are close to the TS condition and special phenomena are expected to show up that can be explored in experiment. Moreover, in case that exotic states can be formed by the $S$-wave interaction between $D_s\bar{D}^*+c.c.$ (and/or $D_s^*\bar{D}+c.c.$) meson pairs, nontrivial lineshape in the invariant mass spectrum of $J/\psi K$ ($J/\psi\bar{K}$) is also expected.

In Ref.~\cite{Voloshin:2018heq} it was proposed that the $S$-wave open threshold $D_{s1}\bar{D}_{s}$ and $D_{s0}\bar{D}^*_{s}$ would feed down to the non-strange open charm decay channel via kaon exchange which would disfavor the formation of a hadronic molecule composed of $D_{s1}\bar{D}_{s}$ and $D_{s0}\bar{D}^*_{s}$~\footnote{This comment was on our earlier version submitted to arXiv, i.e. arXiv:1711.07309}. This is an interesting possibility and for which further discussions will be given later along with the calculations.

As follows, we first present the formalism for the dynamically generated states due to the strong $S$-wave couplings to $D_{s1}\bar{D}_{s}$ and $D_{s0}\bar{D}^*_{s}$ in Section~\ref{sec:dgs}. We then analyze the kinematics of the triangle loops in $e^+e^-\to J/\psi K\bar{K}$ in Section~\ref{sec:ce}. Our calculation results will be compared with the available experimental data, and discussions and conclusions will be presented. A brief summary will be given in the last Section.

\section{Dynamically generated states}\label{sec:dgs}
\begin{figure}[t] \vspace{0.cm}
\begin{center}
\includegraphics[scale=0.5]{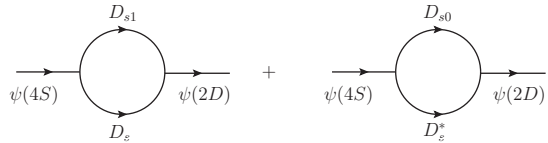}
\caption{Mixing diagrams between $\psi(4S)$ and $\psi(2D)$ via intermediate charmed meson loops.}
\label{fig:pb}
\end{center}
\end{figure}
The mass thresholds for both $D_{s1}\bar D_{s}$ and $D_{s0}\bar D^*_{s}$ lie at about 4.43 GeV (which are 4.428 GeV and 4.429 GeV, respectively), implying the nearly equal spin splitting of mass in the $(1/2)^+$ and $(1/2)^-$ doublets which also happens in the beauty-strange excited meson pairs~\cite{Voloshin:2017mvz}. The $\psi(4S)$ in the potential quark  model with the mass close to these two thresholds can couple to them via an $S$-wave interaction. Given sufficiently strong couplings, it may dynamically generate state near these thresholds by unitarization and result in mixings between the quark model state and the dynamically generated state through the intermediate $D_{s1}\bar D_{s}$ and $D_{s0}\bar D^*_{s}$ bubbles as shown in Fig.~\ref{fig:pb}. Mixings between the $\psi(4S)$ and $\psi(2D)$ via their couplings to these two channels are also considered.

To investigate such a possible scenario, we construct the propagators of $\psi(4S)$ and $\psi(2D)$ in a coupled-channel approach~\cite{Wu:2007jh} as the following:
\ba
G&=&\fr{1}{D_1 D_2-|D_{12}|^2}\begin{pmatrix}D_1&D_{12}\\D_{21}&D_2\end{pmatrix}
\\&=&\fr{G_{12}}{\mr{det}[G_{12}]},
\ea
where $D_1$ and $D_2$ are the denominators of the single propagator of $\psi(4S)$ and $\psi(2D)$, respectively, and $D_{12}$ is the mixing term between them through the $D_{s1}\bar D_{s}$ and $D_{s0}\bar D^*_{s}$ bubble diagrams. So here we have
\ba
D_1&=&m_a^2-s-iB_{11},
\\D_2&=&m_b^2-s-iB_{22},
\\D_{12}&=&iB_{12},
\ea
where $B$ is the sum of the two amplitudes of the bubble diagrams of $D_{s1}\bar D_{s}$ and $D_{s0}\bar D^*_{s}$ between two states. Since $\psi(4S)$ and $\psi(2D)$ both couple to $D_{s1}\bar D_{s}$ and $D_{s0}\bar D^*_{s}$ in an $S$ wave, we have
\ba
B_{11}&=&2g_1^2(I_{20}(P,m_{D_{s1}},m_{D_{s}})+I_{20}(P,m_{D_{s0}},m_{D^*_{s}})),
\\B_{22}&=&2g_2^2(I_{20}(P,m_{D_{s1}},m_{D_{s}})+I_{20}(P,m_{D_{s0}},m_{D^*_{s}})),
\\B_{12}&=&2g_1 g_2(I_{20}(P,m_{D_{s1}},m_{D_{s}})+I_{20}(P,m_{D_{s0}},m_{D^*_{s}})),
\ea
where $g_1$ ($g_2$) is the bare coupling for $\psi(4S)$ ($\psi(2D)$) couples to these two thresholds, $D_{s1}\bar D_{s}$ or $D_{s0}\bar D^*_{s}$; $I_{20}(P,m_a,m_b)$ is the two-point loop integral with the initial energy $P$ and intermediate particle masses $m_a$ and $m_b$. To remove the divergent part of this integral, we adopted an exponential momentum-dependent form factor $\mr{exp}(-2\vec{l}^2/\Lam^2)$ where $\vec l$ is the momentum of the particles in the loop. Such an exponential form factor has an originality from quark model wavefunctions in the harmonic oscillator basis. For an initial meson $S$-wave decays into a meson pair in the final state, the coupling constant defined at hadronic level can always be written as a Gaussian form given by the quark model wavefunction convolution in the harmonic oscillator basis. The cutoff $\Lambda\simeq 1$ GeV indicates the typical hadron size scale. More details about the integral can be found in Appendix A.

\begin{figure}[t] \vspace{0.cm}
\begin{center}
\includegraphics[scale=0.5]{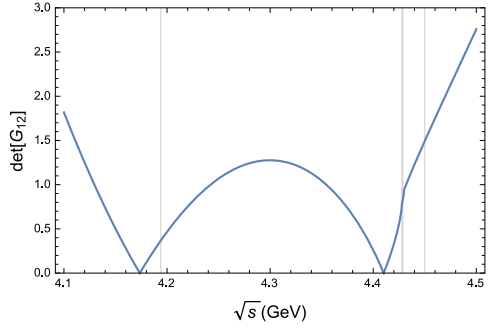}
\caption{Pole structures determined by $\det[G_{12}]=0$ in the propagator matrix. The thick grey vertical line indicates the thresholds of $D_{s1}\bar{D}_s$ and $D_{s0}\bar{D}_s^*$ which are degenerate with each other. The two thin grey lines denote the masses of $\psi(4S)$ and $\psi(2D)$ in the potential quark  model, respectively.}
\label{fig:pdet}
\end{center}
\end{figure}

The corresponding physical states $|A\rangle$ and $|B\rangle$ can be expressed as mixtures of the quark model states $|a\rangle$ and $|b\rangle$ with a mixing matrix, i.e.
\ba\label{mixing-1}
\begin{pmatrix}|A\rangle\\|B\rangle\end{pmatrix}&=&\begin{pmatrix}\mr{cos}\theta&-\mr{sin}\theta e^{i\phi}\\\mr{sin}\theta e^{-i\phi}&\mr{cos}\theta\end{pmatrix}\begin{pmatrix}|a\rangle\\|b\rangle\end{pmatrix}\nonumber\\
&\equiv &R(\theta,\phi)\begin{pmatrix}|a\rangle\\|b\rangle\end{pmatrix},
\ea
With the mixing matrix $R(\theta,\phi)$ the physical propagator matrix $\tilde{G}_{12}$ can be related to $G_{12}$ by
\be
\tilde{G}_{12}=R G_{12}R^{\dagger}.
\ee
The physical propagator matrix $\tilde{G}_{12}$ should be a diagonal matrix. So we can search for the physical poles in the propagator matrix $G$ by requiring $\mr{det}[G_{12}]=0$ after taking into account the $S$-wave open channel effects.

The coupling constants $g_1$ and $g_2$ are unknown parameters in matrix $G_{12}$. In principle, they are not necessarily to be the same because of detailed dynamics, and they can be determined by requiring the physical poles to be located at the masses of the observed states. However, in order to reduce the parameters, we treat $g_1=g_2\equiv g$ for simplicity and as a constraint from the HQSS.  It has been studied in the literature that the $D_{s1}(2460)$ is a mixed state of $^1P_1$ and $^3P_1$ with compatible strength. A mixing scheme similar to Eq.~(\ref{mixing-1}) for the $D_{s1}(2460)$ and $D_{s1}(2536)$ between the quark model $^3P_1$ and $^1P_1$ states has been broadly studied in the literature (see Ref.~\cite{Wu:2011yb} and references therein), i.e.
\begin{eqnarray}\label{Ds-mixing-1}
\begin{pmatrix}|D_{s1}(2460)\rangle\\|D_{s1}(2536)\rangle\end{pmatrix}&=&\begin{pmatrix}\mr{cos}\theta_{D_s}&-\mr{sin}\theta_{D_s} e^{i\phi_{D_s}}\\\mr{sin}\theta_{D_s} e^{-i\phi_{D_s}}&\mr{cos}\theta_{D_s}\end{pmatrix}\begin{pmatrix}|{}^3P_1 \rangle\\|{}^1P_1\rangle\end{pmatrix},\end{eqnarray}
where states $| {}^3P_1 \rangle$ and $| {}^1P_1 \rangle$ can be
rotated to the eigenstates in the heavy quark limit:
\begin{equation}\label{Ds-mixing-2}
    \left(\begin{array}{c}
    | {}^3P_1 \rangle \\ |{}^1P_1 \rangle
    \end{array}\right)=
    \left(\begin{array}{cc}
    \sqrt{\frac{2}{3}} & \sqrt{\frac{1}{3}} \\
    -\sqrt{\frac{1}{3}} & \sqrt{\frac{2}{3}}
    \end{array}\right)
    \left(\begin{array}{c}
    | j=\frac{1}{2} \rangle  \\ | j=\frac{3}{2} \rangle
    \end{array}\right) \ .
\end{equation}
In the above equation $|j=1/2\rangle$ and $|j=3/2\rangle$ denote the total angular momentum of the light quarks in the $D_s$ states, $\vec{j}=\vec{l}+\vec{s}=\vec{1}+\vec{\frac 12}$, which is the commonly used notation for the eigenstates of the heavy-light quark system in the HQSS limit. The advantage of taking the HQSS basis is that one can identify the HQSS favored and suppressed production of the $D_{s1}\bar{D}_{s}$ pairs in $e^+e^-$ annihilations via the corresponding spin decompositions of the meson pair system~\cite{Wang:2013kra}. Combining Eqs.~(\ref{Ds-mixing-1}) and (\ref{Ds-mixing-2}), we have
\begin{eqnarray}
|D_{s1}(2460)\rangle &=& \left[\sqrt{\frac 23}\cos\theta_{D_s}+\sqrt{\frac 13}\sin\theta_{D_s}e^{i\phi_{D_s}}\right]| j=\frac{1}{2} \rangle
+\left[\sqrt{\frac 13}\cos\theta_{D_s}-\sqrt{\frac 23}\sin\theta_{D_s}e^{i\phi_{D_s}}\right]| j=\frac{3}{2}\rangle \ , \\
|D_{s1}(2536)\rangle &=& \left[\sqrt{\frac 23}\sin\theta_{D_s}e^{-i\phi_{D_s}}-\sqrt{\frac 13}\cos\theta_{D_s}\right]| j=\frac{1}{2} \rangle
+\left[\sqrt{\frac 13}\sin\theta_{D_s}e^{-i\phi_{D_s}}+\sqrt{\frac 23}\cos\theta_{D_s}\right]| j=\frac{3}{2}\rangle \ .
\end{eqnarray}
Note that as shown in Ref.~\cite{Wang:2013kra}, the component of $|j=1/2\rangle$ state associated by the $\bar{D}_s$ can couple to $[(c\bar{c})_{1^{--}}(q\bar{q})_{0^{++}}]_{l=0}$, thus, can be produced in $e^+e^-$ annihilations in the HQSS limit. In contrast, the component of $|j=3/2\rangle$ associated by the $\bar{D}_s$ will be forbidden in the HQSS limit. Also, the mixing angles $(\theta_{D_s},\phi_{D_s})= (47.5^\circ, 0^\circ)$ and $(39.7^\circ, (0\pm 6.9)^\circ)$ are determined at the mass of $D_{s1}(2460)$ and $D_{s1}(2536)$, respectively, in Ref.~\cite{Wang:2013kra}. This suggests that the production of the $D_{s1}(2460)\bar D_{s}$ pair is strongly favored than the $D_{s1}(2536)\bar D_{s}$ pair in $e^+e^-$ annihilations near threshold. It is also known that the $S$-wave parts for the $D_{s1}(2460)\bar{D}_s$ and $D_{s0}(2317)D_s^*$ have the same coupling strength in the HQSS~\cite{Wang:2013kra}.

As mentioned earlier, the coupling $g_1$ and $g_2$ are not necessarily to be the same. In particular, in the HQSS limit the $D$-wave charmonium coupling to the $|j=1/2\rangle$ component is forbidden. However, it is also recognized that the HQSS is badly broken in the charmonium system which can be seen in many aspects~\cite{Wang:2013kra,Voloshin:2017awv,Voloshin:2018heq}. A direct experimental evidence is the large decay branching ratio for $\psi(4415)\to D\bar{D}_2^*+c.c.$~\cite{Patrignani:2016xqp} which should occur dominantly via the $D$-wave decays and is consistent with the quark model calculations~\cite{Gui:2018rvv}. This allows that the $D_{s1}\bar{D}_s$ pair can couple to the $S$ and $D$-wave charmonia via the $S$-wave interactions. In this sense, the assumption of $g_1=g_2=g$ is an upper limit for estimating the contributions from the nearby $\psi(2D)$ in this unitarized approach. Whether this is a reasonable assumption can be examined by experimental observables. The first one is the cross section lineshape in the vicinity of these two thresholds. Given the strong $S$-wave interactions with the nearby quark model states, the propagators cannot be described by a simple Breit-Wigner form. Thus, the cross section lineshape will appear to be nontrivial. The second aspect is the decay modes of such mixing states between quark model bare states and dynamically generated states. They will favor decay channels correlated with the threshold interactions. In this case, the reaction channel of $e^+e^-\to J/\psi K\bar{K}$ will be extremely interesting.

Since we still lack experimental data for $e^+e^-\to J/\psi K\bar{K}$ in the vicinity between $\psi(4160)$ and $\psi(4415)$, the following strategy is adopted for investigating the underlying dynamics. By examining the movement of the pole positions of the physical states in terms of coupling $g$ from 1 to $10~\mr{GeV}^{-1/2}$ which is the typical coupling range for the $S$-wave coupling of charmonium-like states to heavy-light $D$ mesons, we identify poles which can match the nearby charmonium states $\psi(4160)$ and $\psi(4415)$ with a reasonable coupling strength for $g$. We find that with $g=7~\mr{GeV}^{-1/2}$ the obtained pole masses by diagonalizing the propagator matrix $G_{12}$ lie at about 4.41 and $4.17~\mr{GeV}$ which are very close to the masses of $\psi(4415)$ and $\psi(4160)$, respectively, as showed in Fig.~\ref{fig:pdet}. It suggests that $\psi(4415)$ as the $\psi(4S)$ charmonium state may contain a sizeable molecular component in its wavefunction caused by its strong couplings to the nearby $S$-wave thresholds, i.e. $D_{s1}\bar{D}_{s}$ and $D_{s0}\bar{D}^*_{s}$. In contrast, the unitarization does not affect much the $\psi(4160)$.

\begin{figure}[t] \vspace{0.cm}
\begin{center}
\includegraphics[scale=0.5]{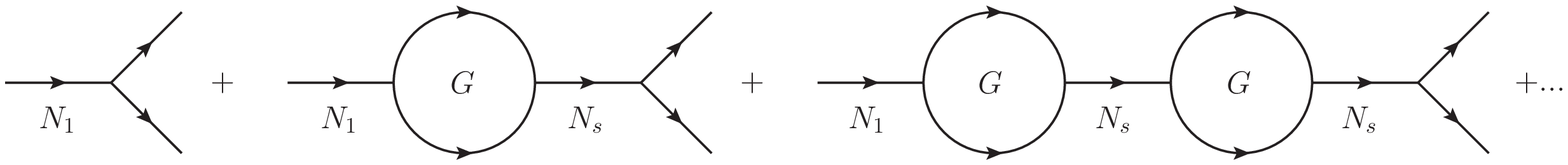}
\caption{The sum of bubbles and bare propagators.}
\label{fig:pp}
\end{center}
\end{figure}

We focus on the energy region near the thresholds of $D_{s1}\bar{D}_{s}$ and $D_{s0}\bar{D}^*_{s}$. For the physical propagator of $\psi(4415)$ we need to sum over all the combinations of bubbles and bare propagators like in Fig.~\ref{fig:pp}. So the physical propagator can be written as
\be
\fr{i}{N_{1p}}=\fr{i}{N_1}G\fr{i}{N_s}+\fr{i}{N_1}G\fr{i}{N_s}G\fr{i}{N_s}+...,
\ee
where $i/N_s=i/N_1+i/N_2$ is the sum of the two bare propagators of $\psi(4S)$ and $\psi(2D)$ and $G(E)$ is the sum of the two kinds of bubbles of $D_{s1}\bar{D}_s$ and $D_{s0}\bar{D}_s^*$ with bare couplings. So we have
\ba
\fr{i}{N_{1p}}&=&\fr{i}{N_1}G\fr{i}{N_s}+\fr{i}{N_1}G\fr{i}{N_s}G\fr{i}{N_s}+...
\\&=&\fr{i}{N_1}/(1-G\fr{i}{N_s})
\\&\equiv &\fr{i}{2m_1(E-m_1-\Sigma_1(E))},
\ea
where $\Sigma_1\equiv iG(E)(N_1+N_2)/(2N_2m_1)$ with $E$ denoting the initial mass energy of $\psi(4415)$ and $m_1$ the bare mass of $\psi(4S)$. By expanding the denominator of the propagator near the physical mass~\cite{Cleven:2013mka,Qin:2016spb} we have
\ba
\fr{2m_1i}{N_{1p}}&=&\fr{i}{E-m_1-\Sigma_1(E)}
\\&=&\fr{iZ}{E-m_1p-Z\widetilde{\Sigma}_1(E)},
\ea
where $\widetilde{\Sigma}_1(E)=\Sigma_1(E)-\mr{Re}[\Sigma_1(m_{1p})]-(E-m_{1p})\mr{Re}[\partial_E{\Sigma_1(m_{1p})}]$ and $Z=1/[1-\mr{Re}(\partial_E{\Sigma_1(m_{1p})})]$ is the wavefunction
renormalization constant, $m_{1p}$ is the physical mass of $\psi(4415)$ and $m_1=m_{1p}-\mr{Re}[\Sigma_1(m_{1p})]$. With $g=7~\mr{GeV}^{-1/2}$ at the pole mass we determine $Z\simeq 0.69$, which suggests that the probability for finding the molecular component in the wavefunction of $\psi(4415)$ is about 50\%, i.e. $(1-Z^2)\simeq 0.52$~\cite{Weinberg:1962hj}. Although this is not a large value for defining a molecular state, we can still examine the consequence arising from such a component in the $\psi(4415)$ wavefunction.

\section{Manifestation of the dynamically generated state in $e^+e^-\to J/\psi K\bar{K}$}\label{sec:ce}

\begin{figure}[t] \vspace{0.cm}
\begin{center}
\includegraphics[scale=0.5]{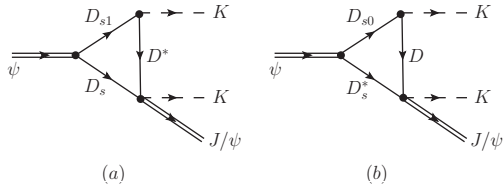}
\caption{The diagrams for $\psi\to J/\psi K\bar K$ via intermediate triangle $D_{(s)}$ meson loops.}
\label{fig:tri}
\end{center}
\end{figure}

With this physical propagator of $\psi(4415)$ and its strong coupling to the $D_{s1}\bar{D}_{s}$ and $D_{s0}\bar{D}^*_{s}$ thresholds we consider the $e^+e^-\to J/\psi K\bar K$ through triangle loops shown in Fig.~\ref{fig:tri}(a) and (b). In these diagrams, the $D_{s1}$ decays into $D^*K$ and $D_{s0}$ decays into $DK$ both in a relative $S$ wave. Also, the scatterings of $\bar{D}_s D^*$ and $\bar{D}_s^*D$ to $J/\psi\bar K$ are also via an $S$ wave. As pointed out earlier, the triangle transition is located in the vicinity of the kinematic condition for the TS. Therefore, it is necessary to investigate the kinematic effects arising from the TS mechanism and identify the dynamically generated states via the $S$-wave interactions with the nearby open thresholds.

To proceed, we first give the corresponding Lagrangians for the $S$-wave coupling:
\ba\label{lagrangian-exp}
\Lag_{\Psi SH}&=&g\langle\Psi \bar{S}^\dag_a\bar{H}_a+\Psi\bar{H}^\dag_a\bar{S}_a\rangle, \nonumber\\
\Lag_{SH\Amp}&=&ih\langle\bar H_a S_b\gam_\mu\gam_5\Amp^\mu_{ba}\rangle, \nonumber\\
\Lag_{HH\Psi\Amp}&=&C\langle \Psi\bar H^\dag_b\gam_\mu\gam_5\bar H_a\Amp^\mu_{ba}\rangle,
\ea
where $S$ and $H$ represent the positive and negative-parity charmed mesons, respectively; while $\Psi$ is the field of the vector charmonium states and $\Amp$ is the chiral field. The explicit form of each field can be found in Appendix B. As mentioned earlier, $g=7~\mr{GeV}^{-1/2}$ is extracted at the pole mass for the $\psi(4415)$. However, the couplings, $h$ and $C$ cannot be determined individually at this moment. As we will show later, by matching the cross sections for $e^+e^-\to J/\psi K\bar{K}$ the product of $h$ and $C$ can be constrained, which shows that these couplings are  within the natural scale, i.e. $\sim O(1)$. 

Typical triangle diagrams for charmonium decaying into $J/\psi K\bar{K}$ are plotted in Fig.~\ref{fig:tri}. For Fig.~\ref{fig:tri} (a), the intermediate $D_{s1}$ plays a key role since it has a strong $S$-wave coupling to $D^*K$. As broadly studied in the literature (see e.g. Ref.~\cite{Guo:2017jvc} and references therein for a recent review of hadronic molecules), the $D_{s1}(2460)$ has been an ideal candidate for a $D^* K$ molecule. Although the mass of $D_{s1}$ is slightly lower than the mass threshold of $D^*K$ which is 2.55 GeV, it has approached the TS kinematics closely with the initial energy also approaching the $D_{s1}\bar{D}_s$ threshold. The presence of the TS also indicates that the dominant contributions from the triangle loop come from the kinematic region where all the internal states are approaching their on-shell condition simultaneously.

One feature arising from the specific process under discussion is that the physical kinematic region for the TS is quite limited. As analyzed in Ref.~\cite{Liu:2015taa}, the physical region for the TS is related to the phase space of the intermediate state two-body decay, i.e. $D_{s1}\to D^* K$. Since the mass of $D_{s1}$ is slightly lower than the $D^*K$ threshold, the contributions from the TS mechanism will be limited to a rather narrow kinematic region. But still, an abnormal lineshape can be expected.

Similar phenomenon happens with the $D_{s0}\bar{D}_s^*(D)$ loop of Fig.~\ref{fig:tri} (b). Also, it should be mentioned that $D_{s0}(2317)$ is an ideal candidate for the $DK$ molecule (see Ref.~\cite{Guo:2017jvc} for a detailed review). Because of the lack of phase space for $D_{s0}\to DK$, the TS kinematics will be restricted within a narrow physical region. But still, observable effects can be expected.

\begin{figure}[t] \vspace{0.cm}
\begin{center}
\includegraphics[scale=0.5]{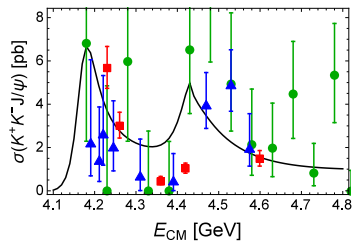}
\caption{The cross section for $e^+e^-\to J/\psi K\bar K$ with the physical propagator of $\psi(4415)$ and via the intermediate triangle $D_{(s)}$ meson loops of Fig.~\ref{fig:tri}. The experimental data are from the Belle~\cite{Yuan:2007bt} (green round dots) and BESIII Collaboration~\cite{Ablikim:2018epj} (blue triangle and red square dots), respectively.}
\label{fig:p3}
\end{center}
\end{figure}

The explicit amplitudes of the diagrams in Fig.~\ref{fig:tri} can be found in Appendix B. Before we come to the calculation results for the final states invariant mass spectra,
we first examine the cross section lineshape for $e^+e^-\to J/\psi K\bar{K}$ around the mass of $\psi(4415)$. In Fig.~\ref{fig:p3} the calculated cross sections are compared with the experimental data from the Belle~\cite{Yuan:2007bt} (green round dots) and BESIII~\cite{Ablikim:2018epj} (blue triangle and red square dots) Collaboration, repsectively. The $\psi(4415)$ as the dynamically generated state which mixes with the quark model state has a lineshape which is apparently deviated from the symmetric Breit-Wigner distribution. It seems that the present data quality still cannot draw a conclusion and such an effect can be investigated with high statistics at BESIII or future Belle-II.

\begin{figure}[t] \vspace{0.cm}
\begin{center}
\includegraphics[scale=0.6]{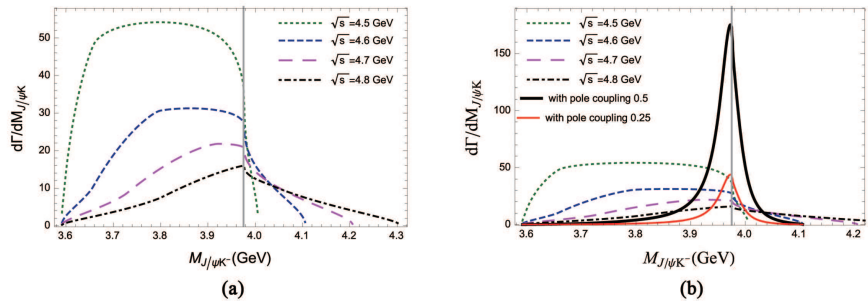}
\caption{The $J/\psi K$ invariant mass spectrum in $\psi\to J/\psi K\bar K$ via (a) intermediate triangle $D_{(s)}$ meson loops, and (b) its comparison with a physical pole. The dotted, dashed, long-dashed and dot-dashed line represent the lineshape at the initial energy of 4.5, 4.6, 4.7 and 4.8 GeV, respectively. The thin and broad solid lines in (b) show the lineshapes with an added pole in the $J/\psi K$ final state which couples to $J/\psi K$ with a natural-scale coupling 0.25 and 0.5, respectively. The grey vertical line indicates the thresholds of $D_{s1}\bar{D}_s$ and $D_{s0}\bar{D}_s^*$  }
\label{fig:ims}
\end{center}
\end{figure}

\begin{figure}[t] \vspace{0.cm}
\begin{center}
\includegraphics[scale=0.6]{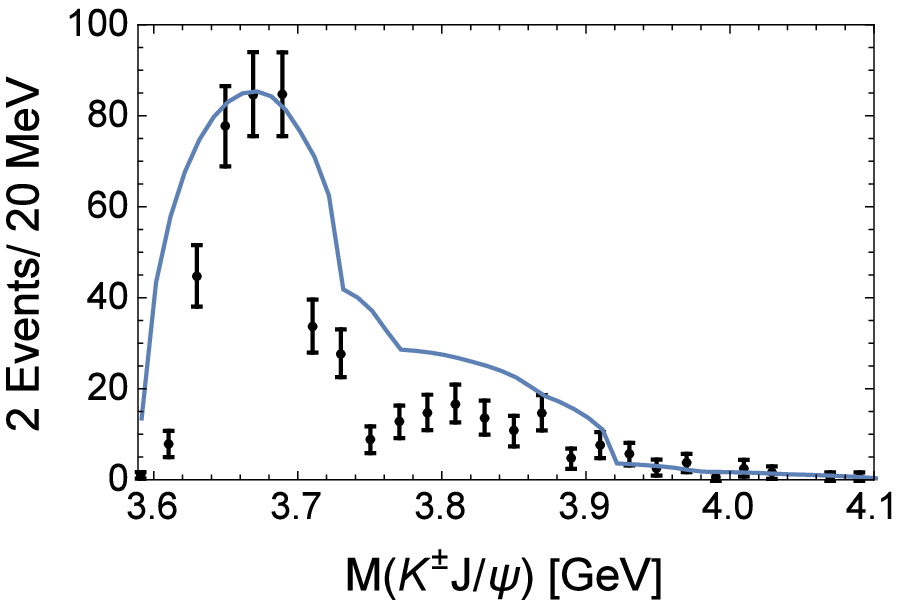}
\caption{The summed $J/\psi K$ invariant mass spectrum in $e^+e^-\to J/\psi K\bar K$ at different c.m. energies the same as the BESIII experiment~\cite{Ablikim:2018epj}, i.e. the cross sections were measured at 14 energy points from $\sqrt{s}=4.189$ to 4.600 GeV. The lineshape appears to be consistent with the data which presumably suggests the absence of $Z_{cs}$ in the $J/\psi K$ spectrum. The vertical axis has an arbitrary unit in order to compare with the experimental events.  }
\label{fig:combined}
\end{center}
\end{figure}

The invariant mass spectrum of the $J/\psi K$ is generally sensitive to the TS mechanism. We plot the $J/\psi K$ spectra in Fig.~\ref{fig:ims} (a) where contributions from Fig.~\ref{fig:tri} (a) and (b) are both included. Also, in order to see the evolution of the TS contributions in terms of the initial energy, we plot the spectra at several energy points from 4.5 to 4.8 GeV. One can see that a CUSP structure, which is located at the common thresholds of $\bar{D}_sD^*$ and $\bar{D}_s^*D$, appears in the $J/\psi K$ invariant mass spectrum. It is difficult to find very clear pole-like structure when the initial mass energy $\sqrt{s}$ is just above the thresholds of $D_{s1}\bar D_s$ or $D_{s0}\bar D_s^*$. But as the initial energy of $\psi(4415)$ increases from 4.5 GeV to 4.8 GeV, a peak-like structure near the threshold of $\bar{D}_sD^*$($\bar{D}_s^*D$) indeed becomes more obvious. Since the mass of $D_{s1}$ is so close to the threshold of $D^*K$, we also discuss the behavior of the spectrum when the mass of $D_{s1}$ is shifted slightly in Appendix C. In Ref.~\cite{Guo:2014iya} it has been shown that lower order singularities than the TS would not produce narrow and pronounced peaks if the interactions between the rescattering hadrons are not strong enough. Similar phenomenon is observed here as shown by Fig.~\ref{fig:ims}. Because of the limited phase space, the TS condition cannot be fully satisfied, thus, the nontrivial threshold structure appears as a CUSP effect instead of the typical narrow peak~\footnote{This scenario will be demonstrated in Appendix C.}. Thus, this process can serve as an ideal channel for the search for possible exotic candidates without ambiguities from the kinematic effects.

In Fig.~\ref{fig:ims} (b) we show the calculations at the initial energy of 4.6 GeV but including explicitly a physical pole right at the mass of the thresholds of $\bar{D}_sD^*$ and $\bar{D}_s^*D$, $\sim 3.98$ GeV, with a typical width of 50 MeV. The consideration is that if there exists the strange partner $Z_{cs}$ of $Z_c(3900)$ as the hadronic molecules of $\bar{D}_sD^*$ and $\bar{D}_s^*D$, the pole structure near the open charm thresholds will produce different lineshapes compared with the kinematic effects shown in Fig.~\ref{fig:ims} (a). Similar to the treatment of Refs.~\cite{Wang:2013cya,Cleven:2013mka,Qin:2016spb} the pole structure can be dynamically generated by the strong $\bar{D}_sD^*$ and $\bar{D}_s^*D$ interactions. Although the detailed dynamics need elaborate studies and are not going to be discussed here, we note that if any mechanism allows the formation of the exotic state with quark contents of $c\bar{c}q\bar{s}$ ($c\bar{c}s\bar{q}$) in this process, the pole structure will appear explicitly in the $J/\psi K$ ($J/\psi\bar{K}$) invariant mass spectrum as the signature for a genuine state.  The thin and broad solid lines in Fig.~\ref{fig:ims} (b) correspond to the pole coupled to $J/\psi K$ with couplings 0.25 and 0.5 of the nature scale, respectively. Compared with the other lines without the pole structure in the $J/\psi K$ invariant mass spectrum, it shows that the pole contributions and pure TS contributions behave quite differently. In this case, the TS mechanism can produce nontrivial lineshapes, but cannot produce predominant peaks at the threshold of $J/\psi K$. If narrow and sharp-peaking structures are observed in the invariant mass spectrum of $J/\psi K$, they can be confidently assigned as signatures for exotics. Also, note that the asymmetric lineshapes are because of the triangle function which will affect the formation of the exotic $Z_{cs}$ state. In this sense, this channel is ideal for testing the TS mechanism and searching for exotic candidates in $e^+e^-$ annihilations.

As a comparison with the newly published BESIII experimental data~\cite{Ablikim:2018epj} we plot the invariant mass spectrum of $K^\pm J/\psi$ in Fig.~\ref{fig:combined} where the events from several different energies are summed up and the corresponding calculation results are also summed up as shown by the solid line. The similarity between the data and theoretical calculation is obvious. This may suggest that a significant signal from the $Z_{cs}$ is disfavored.

The above analysis may allow us to emphasize the following points which can be further investigated for $\psi(4415)$:
\begin{itemize}
\item The $\psi(4415)$ may contain a molecular component in its wavefunction due to its strong $S$-wave coupling to the $D_{s1}\bar D_s$ and $D_{s0}\bar D_s^*$ threshold. Interesting phenomena may arise from this picture in the $\psi(4415)$ decays. As pointed out by Voloshin in Ref.~\cite{Voloshin:2018heq}, these open thresholds can couple to their non-strange partners via the kaon exchange in the $S$-wave. Thus, their decays via the rescatterings of $D_{s1}\bar D_{s}\to D^*\bar{D}_0$ and $D_{s0}\bar D^*_{s}\to D\bar{D}_1$ should be important. The author then comments that this mechanism makes it unlike that a significant admixture of the $D_{s1}\bar D_s$ and $D_{s0}\bar D_s^*$ pairs to be present in the $\psi(4415)$ state. However, this may not be true. A classical example is the $f_0(980)$ which is an ideal candidate for the $K\bar{K}$ molecule and dominantly decays into $\pi\pi$ via the $s\bar{s}$ annihilations. The width of $f_0(980)$ is actually broad, namely, $40\sim 100$ MeV~\cite{Patrignani:2016xqp}. Therefore, the feed-down of partial width to lower mass decay channels cannot be a criterion for molecular structures. However, the feed-down effects on the cross section lineshapes could be a useful signature. Note that the final state will be $D\bar{D}^*\pi+c.c.$ It is interesting to notice the broad enhancement of the cross section in $e^+e^-\to D\bar{D}^*\pi+c.c.$ around $\psi(4415)$~\cite{Ablikim:2018vxx}. The feed-down contribution from the strong couplings of $\psi(4415)$ to the $D_{s1}\bar D_s$ and $D_{s0}\bar D_s^*$ threshold may explain it in addition to the $D_0(2400)\bar{D}^*+c.c.$ contribution.

\item It is possible that if the $\psi(4S)$ has a relatively small coupling to the $D_{s1}\bar D_s$ and $D_{s0}\bar D_s^*$ threshold, then the physical state $\psi(4415)$ should be dominated by the $\psi(4S)$ component. In such a case one needs to understand the relative coupling strength between the $S$ and $P$-wave open charm decay channels. It should also be noted that contributions from $\psi(4415)$ in $e^+e^-\to D^*\bar{D}^*$ and $D_s^*\bar{D}_s^*$ seem to be needed. In the picture of a dominant molecular component for the $\psi(4415)$, its decays into $D^*\bar{D}^*$ and $D_s^*\bar{D}_s^*$ will be suppressed by the power counting which is similar to the case of $Y(4260)$ decaying into $D^*\bar{D}^*$~\footnote{The power analysis is similar to the case of $Y(4260)\to D^*\bar{D}^*$ as recently studied in Ref.~\cite{Xue:2017xpu} except that the $D_{s1}(2460)\to D^* K$ can have an $S$-wave coupling while in the loop transition for $Y(4260)\to D^*\bar{D}^*$ the $D_1(2420)\to D^*\pi$ coupling is via a $D$ wave. As a result, the loop amplitude of $\psi(4415)\to D^*\bar{D}^*$ via the molecular component scales as $v^0\sim 1$ which is relatively suppressed in comparison with that of $\psi(4415)\to D\bar{D}^*\pi+c.c.$ with the $S$-wave $D\pi$ which scales as $1/v$. Here, $v<<1$ is the typical non-relativistic velocity of the intermediate charmed strange mesons in the loop. Since the $D^*K$ threshold is close to the mass of $D_{s1}(2460)$, the intermediate kaon carries a small three-vector momentum for which we estimate that its propagator scales as $1/v^2$.}. A coherent analysis of e.g. $\psi(4415)\to D_{s1}\bar D_s$ (and $D_{s0}\bar D_s^*$) and $\psi(4415)\to D^*\bar{D}^*$ (and $D_s^*\bar{D}_s^*$) will be helpful for clarifying the ambiguities caused by the lack of experimental data. It is also useful to measure the branching ratio fraction between $\psi(4415)\to D^*\bar{D}^*$ and $\psi(4415)\to D\bar{D}^*\pi+c.c.$ with the $S$-wave $D\pi$ as additional probe for the possible molecular structure in the $\psi(4415)$ wavefunction.

\end{itemize}

\section{Summary}
In this work, we investigate phenomena arising from the possible strong couplings of the degenerate thresholds $D_{s1}\bar D_{s}$ and $D_{s0}\bar D^*_{s}$ which may lead to mixings between the dynamically generated hadronic molecule states and the nearby conventional charmonia $\psi(4S)$ and $\psi(2D)$. We find that such a mechanism may have observable effects on $\psi(4415)$ due to the $D_{s1}\bar D_{s}$ and $D_{s0}\bar D^*_{s}$ molecular component. In contrast, its impact on $\psi(4160)$ is relatively small. With the same coupling of $\psi(4415)$ to $D_{s1}\bar D_{s}$ and $D_{s0}\bar D^*_{s}$ we study the $J/\psi K$ final state invariant mass spectrum in $\psi(4415)\to J/\psi K\bar{K}$. It shows that nontrivial lineshapes can be produced by the molecular nature of $\psi(4415)$ in the invariant mass spectrum of $J/\psi K$ which is strongly affected by the fact that the TS mechanism is at the edge of the physical kinematic region.
This provides an ideal channel for testing the TS mechanism on the one hand, and on the other hand, pinning down the process which is sensitive to the production of exotic states $Z_{cs}$ near heavy flavor thresholds. We claim that any predominant peaking structure in the invariant mass spectrum of $J/\psi K$ should confirm its being a genuine state instead of kinematic effects, while absence of the $D^*\bar{D}_s+c.c.$ and $D\bar{D}_s^*+c.c.$ threshold peaking would imply different dynamics for the $S$-wave $D^*\bar{D}_s+c.c.$ ($D\bar{D}_s^*+c.c.$) interaction from that for the $D^*\bar{D}+c.c.$ More experimental data from BESIII and future Belle-II can help clarify such a phenomenon in the future.

\section*{Acknowledgment}

Useful discussions with C. Hanhart, F.-K. Guo and Q. Wang are acknowledged. This work is supported, in part, by the National Natural Science Foundation of China (Grant Nos. 11425525 and 11521505), DFG and NSFC funds to the Sino-German CRC 110 ``Symmetries and the Emergence of Structure in QCD'' (NSFC Grant No. 11261130311), National Key Basic Research Program of China under Contract No. 2015CB856700.

\begin{appendix}
\section{Two-point loop integral}
The $I_{20}$ in Section II can be written in the form
\ba
I_{20}(P,m_1,m_2)&=&\intlf\fc{\mr{exp}(-2\vec l^2/\Lam^2)}{(l^2-m_1^2+i\eps)[(P-l)^2-m_2^2+i\eps]}.
\ea
The form factor parameter $\Lambda$ is taken to be 1 GeV here corresponding to the typical size of hadrons. This integral can be calculated analytically:
\ba
I_{20}&=&\intlf\fc{\mr{exp}(-2\vec l^2/\Lam^2)}{(l^2-m_1^2+i\eps)[(P-l)^2-m_2^2+i\eps]}\nn\\
&=&\fc{i}{4m_1m_2}\fc{4\pi}{(2\pi)^3}\intlz\fc{l^2\ffn}{P-m_1-m_2-l^2/2\mu_{12}}\nn\\
&=&\fc{i}{4m_1m_2}\lt\{-\fc{\mu\Lam}{(2\pi)^{3/2}}+\fc{\mu
k}{2\pi}e^{-2k^2/\Lam^2}\lt[\mr{erfi}\lt(\fc{\st2k}{\Lam}\rt)-i\rt]\rt\},
\ea
where $k=\st{2\mu(M-m_1-m_2)}$ and
$\mu_{ij}$ is the reduced mass of the intermediate particles which are labeled
as $i$ and $j$. The imaginary error function is defined as
\ba
\mathrm{erfi}(z)&=&\fc{2}{\st\pi}\int^z_0\!e^{t^2}\mr{d}t.
\ea

\section{Triangle diagram amplitude}
The fields in the Lagrangians in Eq.~(\ref{lagrangian-exp}) can be written in the form of
\ba
 H_a  &=& \frac{1+{\rlap{v}/}}{2}(\mathcal{D}_{a\mu}^*\gamma^\mu-\mathcal{D}_a\gamma_5) , \\
  S_a &=& \frac{1+{\rlap{v}/}}{2}(\mathcal{D}_{1a}^{\prime \mu}\gamma_\mu\gamma_5-\mathcal{D}_{0a}^*) ,\\
 \Psi &=& \frac{1+{\rlap{v}/}}{2}(\psi(nS)^\mu \gamma_\mu)
\frac{1-{\rlap{v}/}}{2},
\ea
where $a$ is light flavor index. We can then write the amplitude for both triangle diagrams in Fig.~\ref{fig:tri} as follows in the non-relativistic limit of the heavy mesons
\ba
i\mathcal{M}&=&\tilde{g} \vec\eps_\psi\cd\vec\eps_{J/\psi}E_{1K}E_{2K}\nn\\
&&\times [I^{(0)}(m_{D_{s1}},m_{D_s},m_{D^*},P,m_{K^+},m_{J/\psi K^-})+I^{(0)}(m_{D_{s0}},m_{D^*_s},m_{D},P,m_{K^+},m_{J/\psi K^-})\nn\\
&& +I^{(0)}(m_{D_{s1}},m_{D_s},m_{D^*},P,m_{K^-},m_{J/\psi K^+})+I^{(0)}(m_{D_{s0}},m_{D^*_s},m_{D},P,m_{K^-},m_{J/\psi K^+})],
\ea
where $\tilde{g}$ denotes the product of all the coupling constants from the vertices in the triangle diagram, and $I^{(0)}$ is the triangle diagram integral:
\ba
I^{(0)}(m_1,m_2,m_3,M,M_1,M_2)&=&i\intlf\fc{1}{(l^2-m_1^2+i\eps)[(P-l)^2-m_2^2+i\eps][(l-q)^2-m_3^2
+i\eps]}.
\ea
By defining $s=P^2$, $s_1=m^2_{K^+K^-}$ and $s_2=m^2_{J/\psi K^-}$, we have
\ba
m^2_{J/\psi K^+}=s+m^2_{J/\psi}+2m^2_K-s_1-s_2.
\ea
And $E_{1K}\equiv (s+m_K^2-s_2)/(2\sq s)$ is the energy of the individual Kaon at the $SH\Amp$ vertex, $E_{2K}\equiv (s+m_K^2-s-2m^2_K-m_{J/\psi K}^2+s_1+s_2)/(2\sq s)$ is the energy of the Kaon at the $J/\psi K$ vertex. The amplitude is a function depending on $s$, $s_1$ and $s_2$, with $s$ the initial energy squared. The total cross section can be obtained by integrating over $s_1$ and $s_2$ in their phase spaces, and the invariant mass spectrum of $J/\psi K^-$ can be obtained by integrating over only $s_1$.

\section{Triangle singularity condition}
\begin{figure}[t] \vspace{0.cm}
\begin{center}
\includegraphics[scale=0.6]{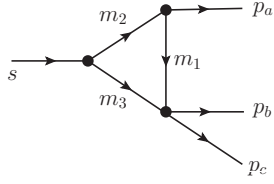}
\caption{Typical 3-body decays via a triangle diagram. Here, we define $s_2=(p_b+p_c)^2$ and $s_3=p_a^2$.}
\label{fig:triexam}
\end{center}
\end{figure}

Early studies of the TS can be found in the literature of last 60's~\cite{Landau:1959fi,Cutkosky:1960sp,bonnevay:1961aa,Eden:1966,barton:1961aa,fronsdal:1964aa,bronzan:1964aa,norton:1964aa,coleman:1965aa}. Its manifestations in high-energy reactions are recognized recently thanks to the high-quality experimental data from B-factories, CLEO-c, BESIII, and LHCb in various exclusive processes. A revival of studying the TS can be found in the recent literature~\cite{Wang:2013cya,Wang:2013hga,Liu:2013vfa,Liu:2015taa,Liu:2015cah,Szczepaniak:2015eza,Guo:2015umn,Wu:2011yx,Guo:2014qra,Bayar:2016ftu}. An up-to-date review can be found in Ref.~\cite{Guo:2017jvc}. Here, we only show how the phase space limits the manifestation of the TS.

For a 3-point loop diagram shown in Fig.~\ref{fig:triexam},  the locations of external momentum variables for different kinds of singularities are determined by the Landau Equation~\cite{Landau:1959fi}. When all the three internal particles get on-shell simultaneously, it pinches the leading singularity of the triangle loop which corresponds to the TS. In Fig.~\ref{fig:triexam}, given the initial energy square $s$, $s_2\equiv (p_b+p_c)^2$, and $s_3\equiv p_a^2$, the locations of the TS can be determined by solving the Landau Equation:
\ba
s^{\pm}&=&(m_2+m_3)^2+\frac{1}{2m_1^2} {\LARGE[}(m_1^2+m_2^2-s_3)(s_2-m_1^2-m_3^2)-4m_1^2 m_2 m_3 \nonumber \\
&\pm& \lambda^{1/2}(s_2, m_1^2, m_3^2)\lambda^{1/2}(s_3,m_1^2,m_2^2){\LARGE ]},
\ea
with $\lambda(x,y,z)= (x-y-z)^2- 4yz$. This is the solutions of $s$ when we fix the masses of all internal particles and $s_2, s_3$. By exchanging $s_1$ and $s_2$, we can obtain the similar solutions for the TS in $s_2$, i.e.,
\begin{eqnarray}
s_2^{\pm}&=&(m_1+m_3)^2+\frac{1}{2m_2^2} {\LARGE[}(m_1^2+m_2^2-s_3)(s-m_2^2-m_3^2)-4m_2^2 m_1 m_3 \nonumber \\  &\pm& \lambda^{1/2}(s,  m_2^2,  m_3^2)\lambda^{1/2}(s_3,m_1^2,m_2^2){\LARGE ]}.
\end{eqnarray}
With the help of the single dispersion relation for the 3-point function, we learn that only $s^-$ or $s_2^-$ corresponds to the TS solutions within the physical boundary~\cite{Liu:2015taa}. The normal and singular thresholds for $s$ and $s_2$ with $s_3$ fixed can be determined as
\ba
&& s_{N}=(m_2+m_3)^2,\ s_{C}=(m_2+m_3)^2 +\frac{m_3}{m_1}[(m_2-m_1)^2-s_3], \nonumber \\
&& s_{2N}=(m_1+m_3)^2,\ s_{2C}=(m_1+m_3)^2 +\frac{m_3}{m_2}[(m_2-m_1)^2-s_3].\label{eq:tsr}
\ea
It describes the motion of the singular thresholds of the TS on the complex plane. Namely, with the fixed $s_3$ and
internal masses, when $s$ reaches $s_{N}$, $s_2^-$ will access its critical threshold $s_{2C}$. Then, with the increase of $s$ from $s_{N}$ to $s_{C}$, $s_2^-$ will move from $s_{2C}$ to $s_{2N}$. This motion will pinch the singularity in the denominator of the dispersion relation, and the range of the motion reflects how significant the TS mechanism can contribute to the loop function. As discussed in Ref.~\cite{Liu:2015taa}, the phase space of internal particle $m_2$ decays into $m_a+m_1$ is correlated with the magnitude of the TS. In our case one notices that $m_2\simeq m_a+m_1$, which means that the TS will be suppressed, or the TS contribution will reduce to a lower order singularity similar to that arising from a two-body cut, i.e. a CUSP effect.

\begin{figure}[t] \vspace{0.cm}
\begin{center}
\includegraphics[scale=0.7]{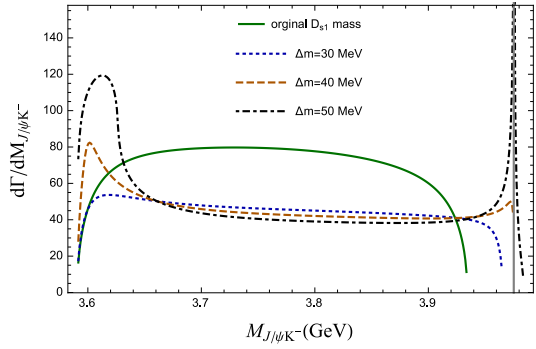}
\caption{The invariant mass spectrum of $J/\psi K^-$ calculated for Fig.~\ref{fig:tri} (a) at the initial energy $\sq s$ near the three body on-shell condition with $m_2=\Delta m+ m_{D_{s1}}$ and $\Delta m=0, 30, 40, 50$ MeV for the solid, dotted, dashed and dot-dashed curve, respectively.}
\label{fig:pts}
\end{center}
\end{figure}

To demonstrate this we plot the invariant mass spectrum for $J/\psi K$  in Fig.~\ref{fig:pts} at the normal threshold of the initial energy $s=(m_{D_{s1}}+m_{D_s})^2$ for the contribution from Fig.~\ref{fig:tri} (a) in $e^+e^-\to J/\psi K\bar{K}$. In order to fulfill the TS condition, we increase the $D_{s1}$ mass $m_{D_{s1}}$  by $\Delta m$ to make it approach to the critical threshold for $s_{2C}$, i.e. to satisfy the on-shell condition for the $D_s\bar{D}^*$ threshold. As shown by the solid curve in Fig.~\ref{fig:pts}, the kinematics for the TS cannot be fulfilled since the mass of the $D_{s1}$ is about 41 MeV below the threshold of $D^*K$. Therefore, the critical threshold $s_{2C}$ does not show up in the invariant mass spectrum. By increasing the mass of $D_{s1}$ by $\Delta m$, the TS will move to the physical kinematic region and the TS effects become more and more important. As shown by the dot-dashed curve in Fig.~\ref{fig:pts} with $\Delta m=$50 GeV, the TS can produce narrow strong peak at the vicinity of the $D_s\bar{D}^*$ threshold. This is a direct demonstration of the role play by the TS when the kinematics are close to the TS condition. It also shows that for the physical case under discussion the observation of predominant peaking structure in the invariant mass spectrum of $J/\psi K$ would imply the existence of a genuine threshold state produced via the triangle transition process.

\end{appendix}

\end{document}